\documentclass[twocolumn]{aastex63}

\usepackage{amsmath}

\newcommand{\nuc}[2]{$^{#1}$#2}
\newcommand{\air}{AIR}

\shorttitle{$r$-process astromers}
\shortauthors{Misch et al.}

\graphicspath{{./}{figs/}}

\begin{document}

\hspace{5.2in} \mbox{LA-UR-20-29246}

\title{Astromers in the radioactive decay of $r$-process nuclei}

\correspondingauthor{G. Wendell Misch}
\email{wendell@lanl.gov}

\author[0000-0002-0637-0753]{G. Wendell Misch}
\affiliation{Theoretical Division, Los Alamos National Laboratory, Los Alamos, NM, 87545, USA}
\affiliation{Center for Theoretical Astrophysics, Los Alamos National Laboratory, Los Alamos, NM, 87545, USA}
\affiliation{Joint Institute for Nuclear Astrophysics - Center for the Evolution of the Elements, USA}

\author[0000-0002-4375-4369]{T.~M. Sprouse}
\affiliation{Theoretical Division, Los Alamos National Laboratory, Los Alamos, NM, 87545, USA}

\author[0000-0002-9950-9688]{M.~R. Mumpower}
\affiliation{Theoretical Division, Los Alamos National Laboratory, Los Alamos, NM, 87545, USA}
\affiliation{Center for Theoretical Astrophysics, Los Alamos National Laboratory, Los Alamos, NM, 87545, USA}
\affiliation{Joint Institute for Nuclear Astrophysics - Center for the Evolution of the Elements, USA}

\begin{abstract}
We study the impact of astrophysically relevant nuclear isomers (astromers) in the context of the rapid neutron capture process ($r$-process) nucleosynthesis.  We compute thermally mediated transition rates between long-lived isomers and the corresponding ground states in neutron-rich nuclei.  We calculate the temperature-dependent $\beta$-decay feeding factors which represent the fraction of material going to each of the isomer and ground state daughter species from the $\beta$-decay parent species.  We simulate nucleosynthesis by including as separate species nuclear excited states with measured terrestrial half-lives greater than 100 $\mu$s.  We find a variety of isomers throughout the chart of nuclides are populated, and we identify those most likely to be influential.  We comment on the capacity of isomer production to alter radioactive heating in an $r$-process environment. 
\end{abstract}

%% Keywords should appear after the \end{abstract} command. 
%% See the online documentation for the full list of available subject
%% keywords and the rules for their use.

\section{Introduction}
\label{sec:intro}

The existence of long-lived excited (metastable) states of atomic nuclei, known as nuclear isomers, was proposed by \cite{soddy1917isomer} and verified by \cite{hahn1921exp} a century ago \citep{walker2020review}.  The study of nuclear isomers has since been a focus of experimental efforts throughout the chart of nuclides, often requiring challenging measurements \citep{raut2013cs, watanabe2014mono, simpson2014yrast, patel2014midshell, svirikhin2017spf, andreev2019isomer, sikorsky2020th}. Over 700 nuclei have been identified as possessing isomeric states with half-lives greater than 100 $\mu$s \citep{ENDFB8}.  Theoretical nuclear structure methods \citep{brown2014shell} are able to characterize some of these isomers, which can arise due to large spin differences (spin traps), differences in nuclear deformation (shape coexistence/shape isomers), and large differences in the projection of spin along the axis of symmetry in a deformed nucleus ($K$ isomers) \citep{dracoulis2016review}. 

Despite the experimental and theoretical progress in terrestrial studies, much remains uncertain regarding the population of nuclear isomers in astrophysical environments \citep{aprahamian2005isomer, hayakawa2005sp, hayakawa2009ncap}.  The most well studied cases in nuclear astrophysics are in relatively light nuclei.  For instance, \nuc{26}{Al} can be used as a tracer of star formation due to the long half-life of the ground state \citep{mahoney1982diffuse, diehl1995comptel}, but its in-situ production is complicated by a low-lying isomer \citep{gupta2001internal, runkle2001thermal, banerjee2018effective}.  The production of \nuc{34}{Cl} may be observable immediately after a nova \citep{coc1999lt}, while \nuc{85}{Kr} is a branch point in the slow neutron capture ($s$) process \citep{abia200185kr} with implications for cosmochronmetry \citep{ward1977importance}; both of these nuclei are influenced by isomers. 

In stark contrast, while mechanisms for inclusion of these additional states in nucleosynthesis codes exist \citep{reifarth2018treatment}, the population of astrophysically relevant nuclear isomers (astromers) is a missing component in the simulation of $r$-process nucleosynthesis.  Exploration of astromers should impact searches for the most influential nuclei to measure at radioactive beam facilities \citep{mumpower2016review, horowitz2019review}, and it overlaps with modern multi-messenger astronomy \citep{abbott2017a, tanvir2017obs, troja2017xray, wang2020gam}. 

In environments where the $r$ process may operate, nuclear isomers can be populated thermally then frozen out as the temperature drops rapidly.  They can also be fed by $\beta$-decay of more neutron-rich isotopes, as well as through other nuclear reactions such as neutron capture and fission \citep{wisshak2006fast, okumura2018fis}.  The direct population of nuclear isomers has recently been shown to be potentially influential in the radioactive heating of $r$-process events by \cite{fujimoto2020isomers}.  They compared a no-isomer simulation against two which included several hand-picked isomers; both isomer models essentially replaced the ground-state properties of the selected nuclei with the isomer properties.

In this work, we explore for the first time dynamical freeze-out as well as the thermal and $\beta$-decay population mechanisms of nuclear isomers in the $r$ process.  We demonstrate that a range of nuclear isomers are significantly populated in the $r$ process between first peak (mass number, $A\sim80$) and third ($A\sim195$) peak elements.  We introduce an Astromer Importance Rating (\air{}) to identify astromers among our included isomers, and we show the effects of our careful treatment of known nuclear isomers on an $r$-process heating curve.  By dynamically including both nuclear isomers and their corresponding ground states, we account for a previously unaddressed non-equilibrium process in simulations of heavy element nucleosynthesis.

\section{Nuclear isomers \& nucleosynthesis}
\label{sec:model}

We treat isomers with half-lives greater than 100 $\mu$s and ground states as distinct ``long-lived'' species.  We use the method of \cite{misch2020astromers} to calculate the thermally mediated transition rates between long-lived states via intermediate states.  This method assumes the states communicate through a thermal photon bath and does not rely on any approximations for the solution to the linear system of equations.  The temperature-dependent $\beta$-decay rates for these long-lived states is computed using the ensemble formalism of \cite{gupta2001internal}.

To account for the feeding of $\beta$-decay daughter nuclei with isomers, we combine laboratory $\beta$-decay $\log(ft)$ values to specific daughter states with that state's probability to contribute to each daughter ensemble; we use available laboratory $\beta$-decay rates and measured $\beta$ intensities to obtain the $\beta$ feeding when $\log(ft)$ values are unavailable.  This yields temperature-dependent decay rates into each of the daughter ``species.''
When $\beta$ intensities are unavailable in the literature, we assume that all decays go to the daughter ground state.

Our isomer calculations use experimental level energies, half-lives, spin-parity assignments, $\gamma$ intensities, and $\log(ft)$ values/$\beta$ intensities from ENSDF\footnote{From ENSDF database as of June 29th, 2020.  Version available at \url{http://www.nndc.bnl.gov/ensarchivals/}}.  We include all experimental levels if there are fewer than 30 reported; if there are more than 30 levels, we include all levels up to the 10 levels above the highest-lying isomer (minimum 30); the analysis of \cite{coc1999lt} shows that this selection is more than adequate to accurately compute transition rates.  For unmeasured $\gamma$ transitions, we use the Weisskopf approximation \citep{weisskopf1930calc}.

We simulate nucleosynthesis with the Jade nuclear reaction network of \cite{sprouse2021jade}.  Jade uses a matrix exponential solver that enables the inclusion of temperature-dependent effects on nuclear transmutation rates.  This allows us to dynamically track the population and de-population of the long-lived states of individual isotopes as a function of time.  For nuclei with isomers, we use the aforementioned treatment. Otherwise, we take laboratory ground-state-to-ground-state half-lives from evaluated data\footnote{\url{https://www-nds.iaea.org/public/download-endf/ENDF-B-VIII.0/}} \citep{ENDFB8, NuBase2016} and supplement with theoretical predictions for yet-to-be-measured nuclei \citep{moller2019beta}. 

A species in our nucleosynthesis network is referenced by two indices: 1) $i$ maps between a tuple consisting of proton number $Z$ and neutron number $N$, and 2) $j$ indicates the individual long-lived state of the nucleus.  For the sake of this work, we use $g$ and $m$ as values of $j$ to denote the ground state and isomer, respectively.  Our network explicitly includes all isomeric states of each isotope, but for clarity, our arguments are presented as though each isotope has at most one isomer.

We focus this work on the population of nuclear isomers in the $r$ process after $\sim$15 minutes.  In this epoch, all relevant neutron captures have finished, and $\beta$-decay, thermally induced excitation, and radiative/thermally stimulated de-excitation are the primary nuclear reaction channels.  We assume an initial solar-like distribution of $r$-process material \citep{arnould2007review} situated far from stability; the distribution consists of all three major $r$-process abundance peaks \citep{sprouse2021jade} as motivated by recent observations \citep{abbott2017a, watson2019sr}.  We have studied the production of nuclear isomers in a variety of astrophysical conditions; here we take the temperature evolution used in \cite{zhu2018cf}, which is based on a homologous expansion into free space \citep{lippuner2015heating}.  We stop our calculations after $\sim$3,600 days (10 years).

\section{\air{}: Astromer Importance Rating}
\label{sec:astromers}

We define the population ratio $R_{i,m}$ of the isomer species of isotope $i$ as
\begin{equation}
    R_{i,m} = \frac{Y_{i,m}}{\sum\limits_{j} Y_{i,j}} \ .
    \label{eq:R}
\end{equation}
The abundance $Y_{i,j}$ is the number of isotope $i$ in state $j$ per baryon of material.  Thus $R_{i,m}$ is the ratio of the isomer abundance to the total abundance of the isotope.  For the sake of brevity and legibility, we suppress the indices on $R$ from here on. 

\begin{figure*}
    \centering
    \includegraphics[width=\textwidth]{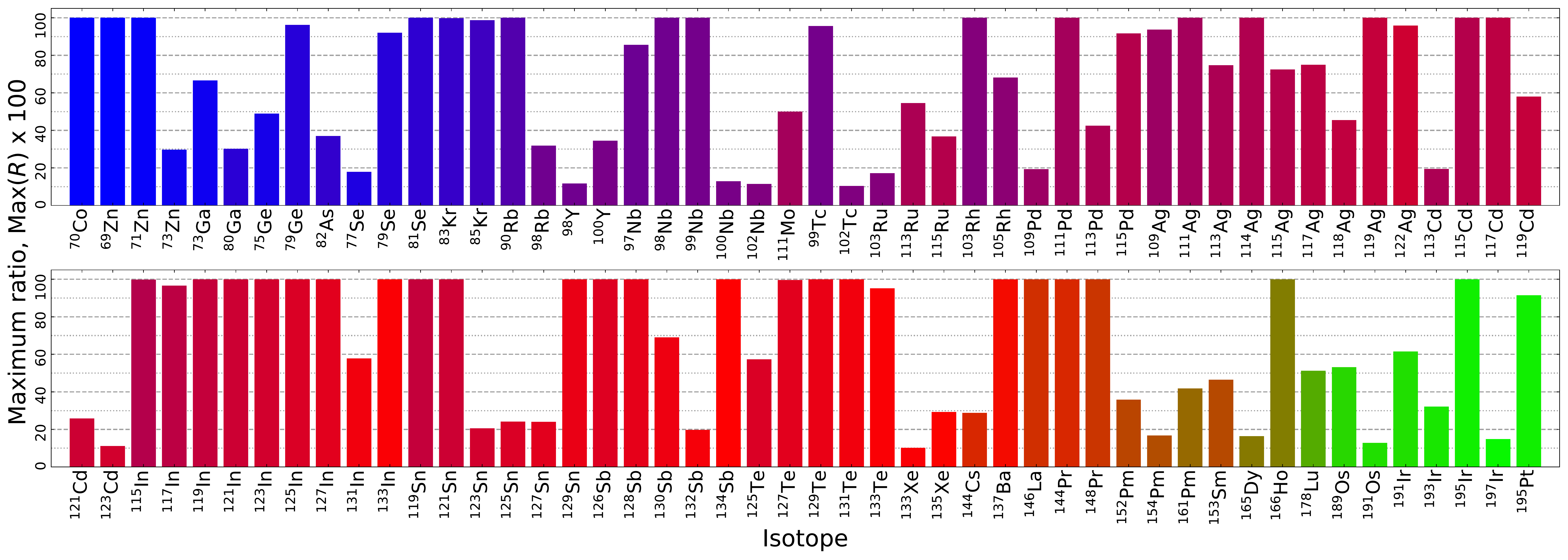}
    \caption{The maximum isomer population ratio, Max($R$), over the course of our simulation for isomers populated above 10\%. Color indicates approximately the $r$-process abundance peak the nucleus resides in; see figure \ref{fig:AIR}.}
    \label{fig:isomer_pop}
\end{figure*}

By finding the maximum value of $R$ over time for each isomer, we can identify which are populated in a nucleosynthesis event.  Figure \ref{fig:isomer_pop} shows this value for the 48 isotopes which at some time in our simulation had $R \ge 0.1$.  Many of the most populated isomers lie in one of the three $r$-process abundance peaks located at $A\sim 80$, $A\sim 130$, and $A\sim 195$. The isomers in these three peaks may be especially significant to the $r$ process, as they not only represent a significant fraction of their respective isotopes, but also these isotopes are among the most abundantly populated during an $r$ process event.

However, not all highly populated isomers are astromers.  They may not freeze out before the isotope has decayed away, or they may not decay at a rate appreciably different from the ground state.  And even if an isomer is an astromer, the nuclear isotope may not be sufficiently abundant to influence $r$-process evolution.  Therefore, we develop a metric to assist in identifying important astromers.

A species affects evolution through its destruction (decay or de-excitation).  For example, consider an abundant ground state of a stable isotope in a cold environment. This species does not participate in the evolution since it neither affects heating nor drives change in abundance.  Similarly, a rapidly decaying isotope with low abundance will not affect the environment.  We calculate the specific activity $a$ of each species to measure its influence. 
\begin{align}
    a_{i,j} &= \lambda_{i,j}Y_{i,j} \\
    \lambda_{i,j} &\equiv \sum\limits_\mathrm{channels} \lambda_{i,j}^\mathrm{channel} \label{eq:lambda}
\end{align}
The $\lambda$s are destruction rates (s$^{-1}$), and the indices $i$ and $j$ are as in equation \ref{eq:R}; we will suppress the $a$ indices $i$ and $j$ going forward, taking $j$ to always be the isomer.  In this work, our channels consist exclusively of $\beta$ decays and thermally-mediated internal transitions.  Activity then quantifies the number of actions taken by an isotope (transitions and/or decays) per baryon of material per second.  The most active isotopes at a given time drive the evolution, provided that that activity is not simply internal transitions back and forth between the ground state and isomer.  This brings us to the thermally weighted abundance imbalance.

% To be an astromer, we also require that the isomer $m$ be destroyed at a rate different from the ground state $g$ (the $\lambda_{i,j}$ in Equation \ref{eq:lambda}); otherwise, it does not meaningfully change the evolution as nucleosynthesis proceeds (compared to $g$ only).  This is especially true at high temperature, where most of the activity could be due to rapid back-and-forth internal transitions.  To quantify how $\lambda_g$ and $\lambda_m$ are different, we use (the absolute value of) their imbalance $I$.

% To be an astromer, we also require that the isomer $m$ behave differently from the ground state $g$; otherwise, it does not meaningfully change the evolution as nucleosynthesis proceeds (compared to $g$ only).  We quantify this using an imbalance  To quantify how $\lambda_g$ and $\lambda_m$ are different, we use (the absolute value of) their imbalance $I$.
% \begin{equation}
%     I(\lambda_g,\lambda_m) = \frac{\vert \lambda_{g} - \lambda_{m} \vert}{\lambda_{g} + \lambda_{m}}
% \end{equation}
% If the destruction rates are equal, the imbalance is zero, and as the rates diverge, the imbalance tends toward unity.  Imbalance has the helpful feature of being independent of the scale of the rates without going to infinity the way a ratio  of the two quantities might; this second point is important when e.g.~the ground state is stable.  Furthermore, unlike a ratio, the imbalance will be large for very different rates regardless of which rate is greater.

To be an astromer, we also require that the isomer~$m$ be out of thermal equilibrium with the ground state~$g$; otherwise, its isomeric quality does not meaningfully impact the evolution as nucleosynthesis proceeds (compared to assuming thermal equilibrium level populations).  To assess how far the populations are from thermal equilibrium, we first define a thermally scaled abundance $\widetilde{Y}$ that enables us to compare the GS and isomer abundances on equal footing.
\begin{equation}
    \widetilde{Y} \equiv \frac{Y}{(2J+1)e^{-E/T}}
\end{equation}
In thermal equilibrium, $\widetilde{Y}_g = \widetilde{Y}_m$.  Now, to quantify how far these values are from equality, we use (the absolute value of) their imbalance $I$.
\begin{equation}
    I(\widetilde{Y}_g,\widetilde{Y}_m) = \frac{\vert \widetilde{Y}_g - \widetilde{Y}_m \vert}{\widetilde{Y}_g + \widetilde{Y}_m}
\end{equation}
If the scaled abundances are equal (in thermal equilibrium), the imbalance is zero, and as they diverge, the imbalance tends toward unity.  Imbalance has the helpful feature of being independent of the scale of the abundances without going to infinity the way a ratio of the two quantities might; this second point is important when e.g.~only one state is populated.  Furthermore, unlike a ratio, the imbalance will be ``large'' (near unity) for inputs which are very different from one another regardless of which is greater.

Taken as a product, the activity $a$, imbalance $I$, and population ratio $R$ give a useful metric---the Astromer Importance Rating \air{}---to help identify populated and influential astromers.
\begin{align}
    \mathrm{AIR} &= \mathrm{Activity} \times \mathrm{Imbalance} \times \mathrm{population~Ratio} \nonumber \\
        &= a \times I \times R
\end{align}

\air{} selects isomers which 1) have a high activity, 2) are far from thermal equilibrium, and 3) have a significant population relative to ground.  These are the necessary ingredients for an influential astromer.  Note that it is possible for a slow-decaying astromer to effectively reduce $a$.  However, at later times, everything with a high $a$ will have decayed away, leaving the slower astromer to dominate the \air{} then.  We therefore search \air{} over a broad range of times to maximize our ability to find important astromers.

\section{Results}\label{sec:results}

\begin{figure*}
    \centering
    \includegraphics[width=\textwidth]{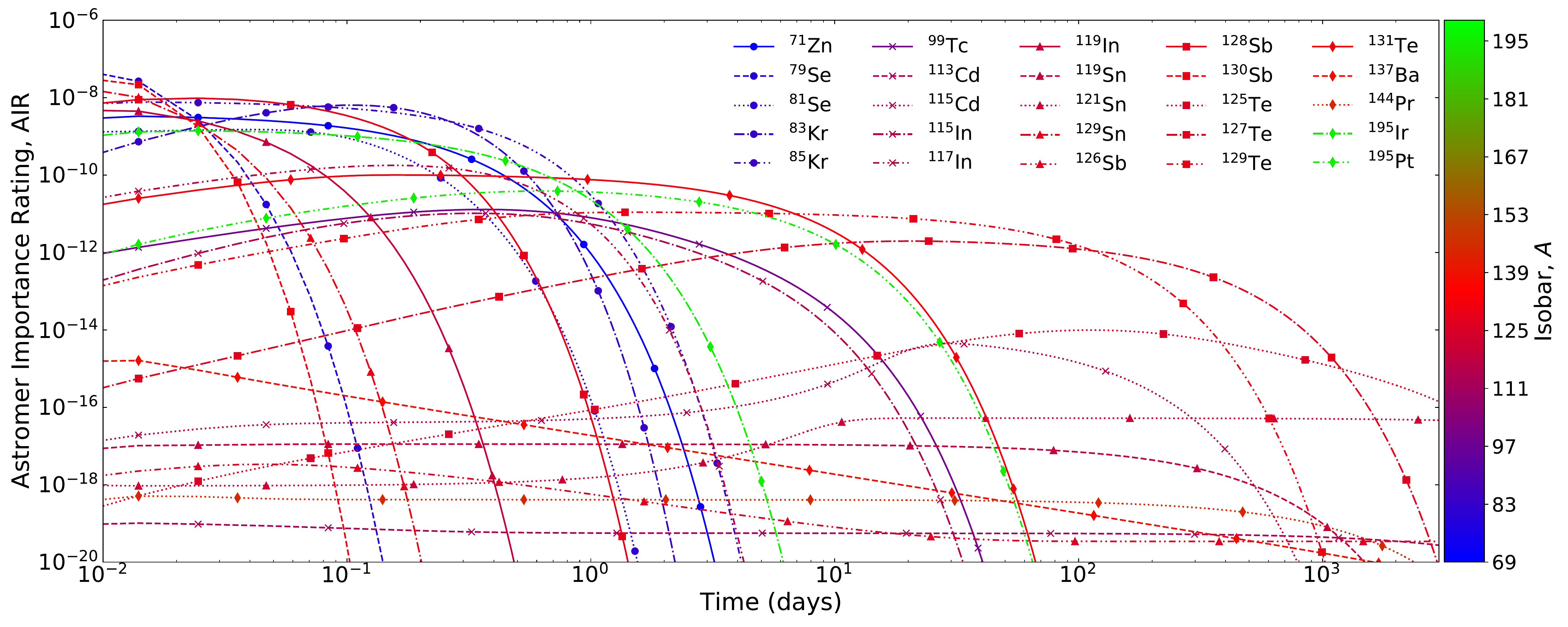}
    \caption{Influential $r$-process astromers ranked by their Astromer Importance Rating, \air{}.  We show isotopes with \air{} values that appear among the top five at some point in our simulation. Color indicates approximately the $r$-process abundance peak the nucleus resides in: 1st (blue), 2nd (red), or 3rd (green).}
    \label{fig:AIR}
\end{figure*}

We computed the \air{} for every nuclear isotope in our simulation to isolate influential astromers.  Figure \ref{fig:AIR} shows the \air{} for the 25 isotopes which, at some simulation timestep, are among the top 5 most important astromers as ranked by \air{}.

Figure \ref{fig:AIR} shows that the most dominant first peak astromers (blue) are relevant for about 1 day with only \nuc{95}{Nb} contributing after 100 days.  A surprising contender on the timescale of a day is \nuc{85}{Kr}, which is most famous as an $s$-process branch point.  In the $r$-process, it has the highest \air{} from $\sim$0.5 to 1 day due to the $>90\%$ feeding of the isomer by \nuc{85}{Br}.  This greatly accelerates the decay to \nuc{85}{Rb}, and it may produce an electromagnetic signal. 

In contrast with the first peak astromers, those located in the second peak (red) persist over much of our simulation, ranging from $0.01$ to $10,000$ days.  As a result, they have the potential to affect many aspects of the evolution throughout a range of observational timescales. At the earliest times in our simulation, around 0.01 days, we note the formation and large \air{} of \nuc{128,130}{Sb} and \nuc{131}{Te}, which contribute alongside a number of additional isomers located in the first and third peaks.  By 30 days, all remaining isomers are located among the second peak nuclei, the most notable of which include \nuc{125,127}{Te}.  Finally, we identify the particularly interesting isomer in \nuc{129}{Te}.  It remains important across all epochs of our simulation, from 0.01 days to 1 year, reflecting a complex interplay between the $\beta$-decay feeding parent, thermal excitation, and de-excitation. 

We identify \nuc{195}{Pt} as the dominant astromer located in the third $r$ process peak (green).  In contrast with the conventional assumption that this isotope is stable upon population via the ground state $\beta$ decay of \nuc{195}{Ir}, our calculations show that the isomer of \nuc{195}{Pt} may be populated directly from the $\beta$ decay of the isomer of \nuc{195}{Ir}.  The \nuc{195}{Pt} isomer de-excites on a timescale of $\sim$4 days, which is relevant to the study of electromagnetic signals associated with the $r$ process. 

Notably, there is one species in Figure \ref{fig:AIR} that does \emph{not} appear in Figure \ref{fig:isomer_pop}: \nuc{95}{Nb}.  This is because while the isomer might not be heavily populated relative to the ground state (low $R$), it has a comparatively high activity $a$ and a large thermal population imbalance $I$.

In the interest of readability, Figure \ref{fig:AIR} shows only those astromers which rank in the top 5 by \air{}. When we expand our enumeration of $r$-process astromers to include the top 10, we identify only 11 additional astromers. This highlights the effectiveness of \air{} as a filter for distinguishing astromers from the broader population of isomers: the list of isomers identified as \textit{astromers} is relatively insensitive to the particular choice of \air{} top-$N$.  We summarize our expanded (top 10) list of astromers in Table \ref{tab:top10}.

% Table 1
\begin{deluxetable*}{ccccccccl}
    \tablecaption{Astromer importance rating (\air{}) top 10 selected at each time point $t\gtrsim 15$ min.  We use $g$ and $m$ to indicate the ground state and isomer, respectively.  The isomer energy is $E_m$, and the $J^\pi$ are the spin and parity of the respective levels (parentheses denote uncertain $J^\pi$).  The half-lives ($T_{1/2}$) and $\beta$-decay branching for the isomer ($B_{m\beta}$) are as measured in the laboratory; $B_{m\beta}$ is the percent of isomer decays which are $\beta$ decays rather than internal transitions to another nuclear state.  $T_{pop}$ is the approximate timescale on which the isotope is populated, and Notes gives some brief comments on the nucleus.}
    
    \tablehead{
    \colhead{Isotope} & \colhead{$E_{m}$} & \colhead{$J^\pi_{g}$} & \colhead{$J^\pi_{m}$} & \colhead{$T_{1/2~g}$} & \colhead{$T_{1/2~m}$} & $B_{m\beta}$   & $T_{pop}$ & Notes \\ 
    \colhead{}        & \colhead{(keV)}   & \colhead{}            & \colhead{}            & \colhead{(s)}         & \colhead{(s)}         & \colhead{(\%)} & \colhead{}  & \colhead{}
}
    
    \startdata
    % \nuc{69}{Ni}  & 321     & $(9/2^{+})$ & $(1/2^{-})$  & $1.14\times 10^{1}$  & $3.5\times 10^{0}$   & 100    & \nodata \\
    % \nuc{70}{Cu}  & 101.1   & $6^{-}$     & $3^{-}$      & $4.45\times 10^{1}$  & $3.3\times 10^{1}$   & 52     & \nodata \\
    \nuc{69}{Zn}  & 438.636 & $1/2^{-}$   & $9/2^{+}$    & $3.38\times 10^{3}$  & $4.95\times 10^{4}$  & 0.033  & 3 min    & $\lambda^\beta$ slowed for 14 hours, EM signal\tablenotemark{a} \\
    \nuc{71}{Zn}  & 157.7   & $1/2^{-}$   & $9/2^{+}$    & $1.47\times 10^{2}$  & $1.43\times 10^{4}$  & 100    & 20 s     & $\lambda^\beta$ slowed for 4 hours \\
    % \nuc{77}{Zn}  & 772.440 & $(7/2^{+})$ & $(1/2^{-})$  & $2.08\times 10^{0}$  & $1.05\times 10^{0}$  & 66     & \nodata \\
    \hline
    % \nuc{79}{Ge}  & 185.95  & $(1/2)^{-}$ & $(7/2^{+})$  & $1.898\times 10^{1}$ & $3.9\times 10^{1}$   & 96     & \nodata \\
    \nuc{79}{Se}  & 95.77   & $7/2^{+}$   & $1/2^{-}$    & $1.03\times 10^{13}$ & $2.35\times 10^{2}$  & 0.056  & 9 min    & $T_{1/2~m} < T_{pop}$, no new effect \\
    % \nuc{80}{Ga}  & 22.4    & $6^{(-)}$   & $3^{(-)}$    & $1.9\times 10^{0}$   & $1.3\times 10^{0}$   & 100    & \nodata \\
    \nuc{81}{Se}  & 103.00  & $1/2^{-}$   & $7/2^{+}$    & $1.11\times 10^{3}$  & $3.44\times 10^{3}$  & 0.051  & 30 s     & $\lambda^\beta$ slowed for 1 hour \\
    \hline
    \nuc{83}{Kr}  & 41.5575 & $9/2^{+}$   & $1/2^{-}$    & stable               & $6.59\times 10^{3}$  & 0      & 2.5 h    & $\gamma$ only, likely no effect \\
    \nuc{85}{Kr}  & 304.871 & $9/2^{+}$   & $1/2^{-}$    & $3.39\times 10^{8}$  & $1.61\times 10^{4}$  & 78.8   & 3 min    & $\lambda^\beta$ accelerated for 5 hr\tablenotemark{b}, EM signal\tablenotemark{a} \\
    \hline
    % \nuc{90}{Rb}  & 106.90  & $0^{-}$     & $3^{-}$      & $1.58\times 10^{2}$  & $2.58\times 10^{2}$  & 97.4   & \nodata \\
    \nuc{93}{Nb}  & 30.77   & $9/2^{+}$   & $1/2^{-}$    & stable               & $5.09\times 10^{8}$  & 0      & 1.6 Myr  & $T_{1/2} < T_{pop}$, no new effect \\
    \nuc{95}{Nb}  & 235.69  & $9/2^{+}$   & $1/2^{-}$    & $3.02\times 10^{6}$  & $3.12\times 10^{5}$  & 5.6    & 64 d     & $T_{1/2} < T_{pop}$, no new effect \\
    \nuc{97}{Nb}  & 743.35  & $9/2^{+}$   & $1/2^{-}$    & $4.33\times 10^{3}$  & $5.87\times 10^{1}$  & 0      & 17 h     & $T_{1/2} < T_{pop}$, no new effect \\
    \hline
    \nuc{99}{Tc}  & 142.684 & $9/2^{+}$   & $1/2^{-}$    & $6.66\times 10^{12}$ & $2.16\times 10^{4}$  & 0.0037 & 66 h     & $T_{1/2~m} < T_{pop}$, no new effect \\
    \hline
    \nuc{113}{Cd} & 263.54  & $1/2^{+}$   & $11/2^{-}$   & $2.54\times 10^{23}$ & $4.45\times 10^{8}$  & 99.86  & 6 h      & low $Y_m$, long $T_{1/2}$, likely unobservable \\
    \nuc{115}{Cd} & 181.0   & $1/2^{+}$   & $(11/2)^{-}$ & $1.92\times 10^{5}$  & $3.85\times 10^{6}$  & 100    & 20 min   & $\lambda^\beta$ slowed for 45 d \\
    \nuc{117}{Cd} & 136.4   & $1/2^{+}$   & $(11/2)^{-}$ & $8.96\times 10^{3}$  & $1.21\times 10^{4}$  & 100    & 70 s   & $\lambda^\beta$ slowed for 3.4 h \\
    \hline
    \nuc{115}{In} & 336.244 & $9/2^{+}$   & $1/2^{-}$    & $1.39\times 10^{22}$ & $1.61\times 10^{4}$  & 5.0    & 54 h     & \nuc{115}{Sn} production boosted, EM signal\tablenotemark{a} \\
    \nuc{117}{In} & 315.303 & $9/2^{+}$   & $1/2^{-}$    & $2.59\times 10^{3}$  & $6.97\times 10^{3}$  & 52.9   & 3 h      & $T_{1/2} < T_{pop}$, no new effect \\
    \nuc{119}{In} & 311.37  & $9/2^{+}$   & $1/2^{-}$    & $1.44\times 10^{2}$  & $1.08\times 10^{3}$  & 95.6   & 3 min    & $\lambda^\beta$ slowed for 18 min \\
    % \nuc{121}{In} & 313.68  & $9/2^{+}$   & $1/2^{-}$    & $2.31\times 10^{1}$  & $2.33\times 10^{2}$  & 98.8   & 15 s     & feeds \nuc{121}{Sn} isomer \\
    \hline
    \nuc{119}{Sn} & 89.531  & $1/2^{+}$   & $11/2^{-}$   & stable               & $2.53\times 10^{7}$  & 0      & 3-18 min & EM signal\tablenotemark{a} \\
    \nuc{121}{Sn} & 6.31    & $3/2^{+}$   & $11/2^{-}$   & $9.73\times 10^{4}$  & $1.39\times 10^{9}$  & 22.4   & 4 min    & $\lambda^\beta$ slowed for 44 yr, EM signal\tablenotemark{a} \\
    \nuc{129}{Sn} & 35.15   & $3/2^{+}$   & $11/2^{-}$   & $1.34\times 10^{2}$  & $4.14\times 10^{2}$  & 100    & 1 s      & $\lambda^\beta$ slowed for 7 min \\
    \hline
    % \nuc{123}{In} & 327.21  & $(9/2)^{+}$ & $(1/2)^{-}$  & $6.17\times 10^{0}$  & $4.74\times 10^{1}$  & 100    & \nodata \\
    % \nuc{125}{In} & 360.12  & $9/2^{+}$   & $1/2^{(-)}$  & $2.36\times 10^{0}$  & $1.22\times 10^{1}$  & 100    & \nodata \\
    \nuc{126}{Sb} & 17.7    & $(8^{-})$   & $(5^{+})$    & $1.07\times 10^{6}$  & $1.15\times 10^{3}$  & 86     & 230 kyr  & $T_{1/2} < T_{pop}$, no new effect \\
    \nuc{128}{Sb} & 0.0+X   & $8^{-}$     & $5^{+}$      & $3.26\times 10^{4}$  & $6.25\times 10^{2}$  & 96.4   & 1 h      & $\lambda^\beta$ accelerated to 11 min $T_{1/2}$\tablenotemark{b} \\
    \nuc{130}{Sb} & 4.8     & $(8^{-})$   & $(4,5)^{+}$  & $2.37\times 10^{3}$  & $3.78\times 10^{2}$  & 100    & 4 min    & $\lambda^\beta$ accelerated to 6.3 min $T_{1/2}$\tablenotemark{b} \\
    \hline
    % \nuc{127}{Sn} & 5.07    & $11/2^{-}$  & $3/2^{+}$    & $7.56\times 10^{3}$  & $2.48\times 10^{2}$  & 100    & \nodata \\
    \nuc{125}{Te} & 144.775 & $1/2^{+}$   & $11/2^{-}$   & stable               & $4.96\times 10^{6}$  & 0      & 3 yr     & $T_{1/2~m} < T_{pop}$, no new effect \\
    \nuc{127}{Te} & 88.23   & $3/2^{+}$   & $11/2^{-}$   & $3.37\times 10^{4}$  & $9.17\times 10^{6}$  & 2.4    & 4 d      & $\lambda^\beta$ slowed for 100 d, EM signal\tablenotemark{a} \\
    % \nuc{128}{In} & 340     & $(3)^{+}$   & $(8^{-})$    & $8.4\times 10^{-1}$  & $7.2\times 10^{-1}$  & 100    & \nodata \\
    \nuc{129}{Te} & 105.51  & $3/2^{+}$   & $11/2^{-}$   & $4.18\times 10^{3}$  & $2.9\times 10^{6}$   & 36     & 4.5 h    & $\lambda^\beta$ slowed for 34 d, EM signal\tablenotemark{a} \\
    % \nuc{131}{In} & 302     & $(9/2^{+})$ & $(1/2^{-})$  & $2.8\times 10^{-1}$  & $3.5\times 10^{-1}$  & 99.982 & \nodata \\
    % \nuc{131}{Sn} & 0.0+X   & $(3/2^{+})$ & $(11/2^{-})$ & $5.6\times 10^{1}$   & $5.84\times 10^{1}$  & 100    & \nodata \\
    \nuc{131}{Te} & 182.258 & $3/2^{+}$   & $11/2^{-}$   & $1.5\times 10^{3}$   & $1.2\times 10^{5}$   & 74.1   & 23 min   & $\lambda^\beta$ slowed for 33 h, EM signal\tablenotemark{a} \\
    \nuc{133}{Te} & 334.26  & $(3/2^{+})$ & $(11/2^{-})$ & $7.5\times 10^{2}$   & $3.32\times 10^{3}$  & 83.5   & 2.5 min  & $\lambda^\beta$ slowed for 1 h \\
    \hline
    \nuc{131}{Xe} & 163.930 & $3/2^{+}$   & $11/2^{-}$   & stable               & $1.02\times 10^{6}$  & 0      & 8 d      & EM signal\tablenotemark{a} \\
    \nuc{133}{Xe} & 233.221 & $3/2^{+}$   & $11/2^{-}$   & $4.53\times 10^{5}$  & $1.9\times 10^{5}$   & 0      & 21 h     & EM signal\tablenotemark{a} \\
    \hline
    \nuc{137}{Ba} & 661.659 & $3/2^{+}$   & $11/2^{-}$   & stable               & $1.53\times 10^{2}$  & 0      & 30 yr    & $T_{1/2} < T_{pop}$, no new effect \\
    \hline
    \nuc{144}{Pr} & 59.03   & $0^{-}$     & $3^{-}$      & $1.04\times 10^{3}$  & $4.32\times 10^{2}$  & 0.07   & 285 d    & $T_{1/2} < T_{pop}$, no new effect \\
    \hline
    \nuc{166}{Ho} & 5.969   & $0^{-}$     & $7^{-}$      & $9.66\times 10^{4}$  & $3.79\times 10^{10}$ & 100    & 82 h     & $\lambda^\beta$ slowed for 1200 y \\
    \hline
    \nuc{189}{Os} & 30.82   & $3/2^{-}$   & $9/2^{-}$    & stable               & $2.09\times 10^{4}$ & 0       & 1 d      & $T_{1/2} < T_{pop}$, no new effect \\
    \hline
    \nuc{191}{Ir} & 171.29  & $3/2^{+}$   & $11/2^{-}$   & stable               & $4.90\times 10^{0}$  & 0      & 16 d     & $T_{1/2~m} < T_{pop}$, no new effect \\
    \nuc{195}{Ir} & 100     & $3/2^{+}$   & $11/2^{-}$   & $8.24\times 10^{3}$  & $1.32\times 10^{4}$  & 95     & 7 min    & feeds \nuc{195}{Pt} isomer\tablenotemark{c} \\
    \hline
    \nuc{195}{Pt} & 259.077 & $1/2^{-}$   & $13/2^{+}$   & stable               & $3.46\times 10^{5}$  & 0      & 4 h      & EM signal\tablenotemark{a} \\
    \enddata
    
    \tablenotetext{a}{EM signals (detectable x-rays and $\gamma$-rays) are \emph{possible}, but we do not study them carefully here.}
    \tablenotetext{b}{Nearly all parent $\beta$ decay feeds the isomer, effectively bypassing the longer-lived ground state.}
    \tablenotetext{c}{To explore \nuc{195}{Pt} isomer population, we assume all \nuc{195}{Os} $\beta$ decay feeds the \nuc{195}{Ir} isomer (actual feeding unknown).}
    
    \label{tab:top10}

\end{deluxetable*}

The astromers identified in Table \ref{tab:top10} that change $\beta$-decay rates will affect $r$-process heating.  Some (e.g. \nuc{115}{Cd}, \nuc{127,129,131}{Te}) will defer heating by delaying decay, while others (e.g. \nuc{85}{Kr}, \nuc{128,130}{Sb}) partially counteract this trend by accelerating decay. 

Furthermore, astromers may generate identifiable x-ray or $\gamma$-ray signals either directly through de-excitation (\nuc{85}{Kr}, \nuc{115}{In}, \nuc{119}{Sn}, \nuc{131,133}{Xe}, \nuc{195}{Pt}) or indirectly subsequent to $\beta$ decay (any with $T_{1/2}\gtrsim 1$ day).  Such isomers could be used to directly associate future $r$ process observations with the production of specific nuclei.

We emphasize two points about the relationship between available nuclear data and our astromer calculations. First, more complete excited-state data would improve our ground state $\leftrightarrow$ isomer transition rates.  Specifically, measurements of intermediate state half-lives and $\gamma$ intensities enable more reliable calculations of thermally mediated transition rates.  This is exemplified by \nuc{128}{Sb}, whose isomer dramatically accelerates the decay of the $A=128$ mass chain.  All measured excited states in this nucleus connect to the isomer---which itself has an unknown energy---and the missing links to ground suppress our calculated transition rate.

Second, we need more complete $\beta$ intensities to compute reliable feeding factors.  Because we assume that decays with unpublished intensities always go to ground in the daughter, we preclude the possibility that such decays populate a daughter isomer.  The production of \nuc{170}{Ho} is one example of this issue: the $\beta$ intensities of \nuc{170}{Dy} are unknown, leading to our inability to effectively calculate the feeding factors into the \nuc{170}{Ho} isomer \citep{misch2020astromers}.  The $A=115$ and $A=129$ mass chains are particularly striking examples of missing $\beta$-decay information.  While the feeding factors from the \nuc{115}{Cd} and \nuc{129}{Sn} parents are measured, we lack data on how multiple earlier ancestors populate isomers in those parents.  Experimental and theoretical efforts to quantify unknown $\beta$ intensities would improve understanding of the population of astromers in $r$-process events.

\begin{figure}
    \centering
    \includegraphics[width=\columnwidth]{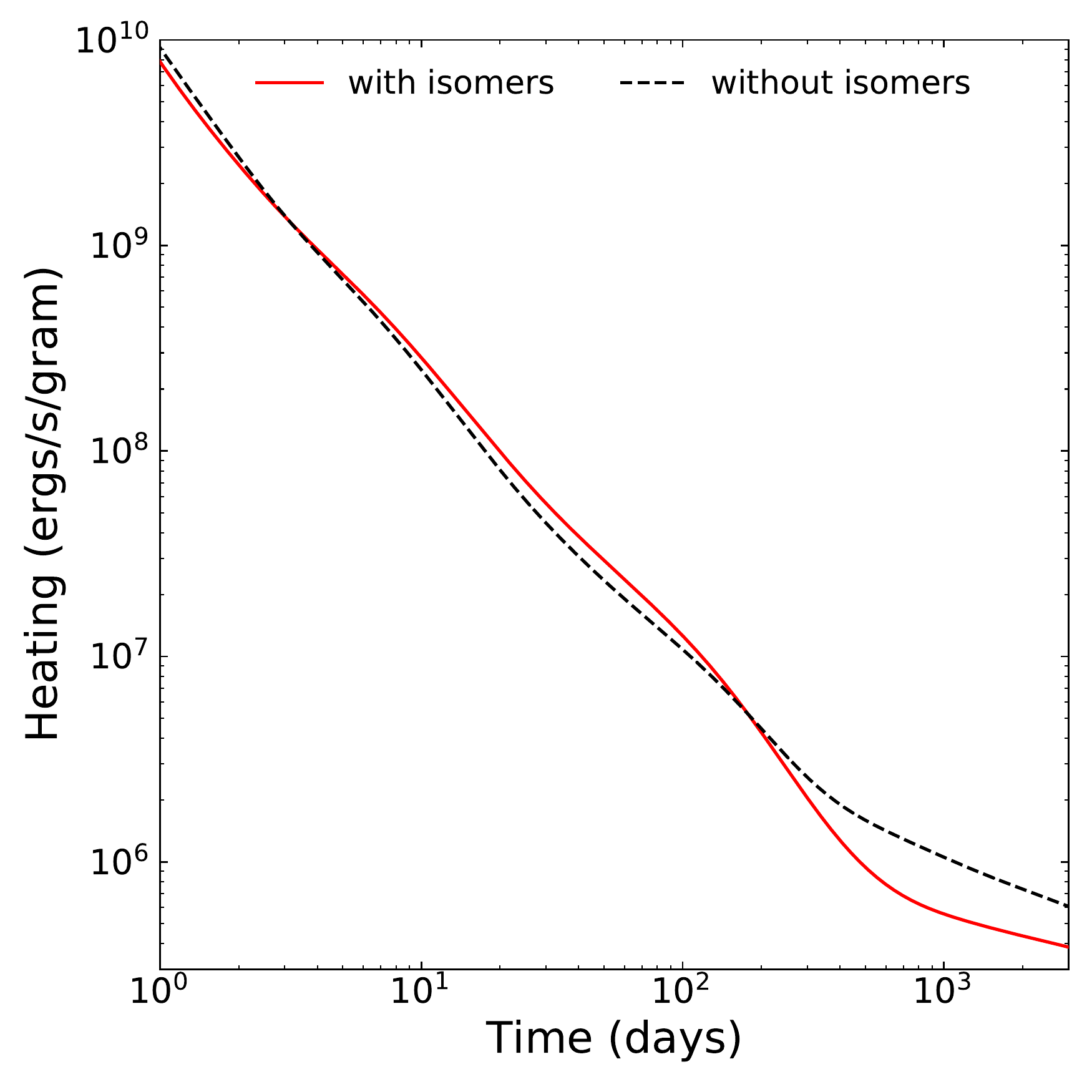}
    \caption{Radioactive heating from $\beta$-decay with and without isomers. We find a 25\% increase in heating from the inclusion of isomers between three and one hundred days.}
    \label{fig:isomer_heating}
\end{figure}

One of the key roles that isomers have recently been proposed to play is in adjusting the timescale on which nuclear energy is released during the radioactive decay towards stability immediately following $r$-process nucleosynthesis \citep{fujimoto2020isomers}.  To investigate the possibility of these effects, we performed calculations for the radioactive heating of $r$-process material both with and without the consideration of isomers.  We show our calculations of these two heating rates as a function of time in Figure \ref{fig:isomer_heating}.  

Within the first day, astromers effectively store energy by slowing radioactive decay towards stability as a result of their (on average) comparatively longer half-lives. On longer timescales (10-100 days), the slower decay of these astromers leads to a gradual release of the energy they had previously retained, with the effect of boosting the heating by about 25\%.  Analogous cycles of energy retention and release by astromers tend to repeat on longer timescales, the next of which begins around 200 days.  We find the inclusion of temperature-dependent $\beta$-decay rates, as well as thermally induced excitation and spontaneous/thermally stimulated de-excitation between long-lived nuclear states, tempers the dramatic effects suggested by \cite{fujimoto2020isomers}.

\section{Conclusions} \label{sec:con}

We have calculated temperature-dependent production and destruction rates ($\beta$ decay and internal transitions) using available data for all known neutron-rich isomers that may participate in the rapid neutron capture process.  We show for the first time that nuclear isomers are dynamically populated in an $r$-process event. 

We focus on those populated isomers of notable astrophysical importance, and refer to these as astromers.  Our nucleosynthesis simulations for the radioactive decay of $r$ process nuclei, in combination with our newly defined Astromer Importance Rating, \air{}, facilitates the identification of influential $r$-process astromers.  

Our work highlights the need for careful treatment of isomeric states when simulating the many and varied aspects of $r$-process nucleosynthesis.  This is especially true when attempting to incorporate astromers that may originate from fission deposition or be dynamically produced and/or destroyed at early times when neutron capture is still active. 

Current \citep{zhang2019isomer, orford2020isomer, nesterenko2020isomer, walker2020properties} and future studies at radioactive beam facilities that focus on elucidating the properties of nuclear isomers  will help to refine our astromer predictions and strengthen our understanding of the creation of the heavy elements.  This work provides fresh motivation for experimental campaigns to measure the beyond-ground-state properties---such as excited state half-lives and $\gamma$-ray intensities---that astromer evolution relies on.

% \section{Acknowledgements}\label{sec:ack}
\acknowledgments

We thank A.~Couture, C.~Fryer, B.~Meyer, and F.~Timmes for valuable discussions.  G.W.M., T.M.S. and M.R.M. were supported by the US Department of Energy through the Los Alamos National Laboratory (LANL). LANL is operated by Triad National Security, LLC, for the National Nuclear Security Administration of U.S.\ Department of Energy (Contract No.\ 89233218CNA000001). 
G.W.M and M.R.M. were partly supported by the Laboratory Directed Research and Development program of LANL under project number 20190021DR. 
T.M.S. was partly supported by the Fission In R-process Elements (FIRE) Topical Collaboration in Nuclear Theory, funded by the U.S. Department of Energy. 

\bibliography{references}

\begin{thebibliography}{}
\expandafter\ifx\csname natexlab\endcsname\relax\def\natexlab#1{#1}\fi
\providecommand{\url}[1]{\href{#1}{#1}}
\providecommand{\dodoi}[1]{doi:~\href{http://doi.org/#1}{\nolinkurl{#1}}}
\providecommand{\doeprint}[1]{\href{http://ascl.net/#1}{\nolinkurl{http://ascl.net/#1}}}
\providecommand{\doarXiv}[1]{\href{https://arxiv.org/abs/#1}{\nolinkurl{https://arxiv.org/abs/#1}}}

\bibitem[{{Abbott} {et~al.}(2017){Abbott}, {Abbott}, {Abbott}, {Acernese},
  {Ackley}, {Adams}, {Adams}, {Addesso}, {Adhikari}, {Adya}, \&
  et~al.}]{abbott2017a}
{Abbott}, B.~P., {Abbott}, R., {Abbott}, T.~D., {et~al.} 2017, \apjl, 848, L12,
  \dodoi{10.3847/2041-8213/aa91c9}

\bibitem[{Abia {et~al.}(2001)Abia, Busso, Gallino, Dom{\'\i}nguez, Straniero,
  \& Isern}]{abia200185kr}
Abia, C., Busso, M., Gallino, R., {et~al.} 2001, The Astrophysical Journal,
  559, 1117

\bibitem[{Andreev {et~al.}(2019)Andreev, Savel'ev, Stremoukhov, \&
  Shoutova}]{andreev2019isomer}
Andreev, A.~V., Savel'ev, A.~B., Stremoukhov, S.~Y., \& Shoutova, O.~A. 2019,
  Phys. Rev. A, 99, 013422, \dodoi{10.1103/PhysRevA.99.013422}

\bibitem[{{Aprahamian} \& {Sun}(2005)}]{aprahamian2005isomer}
{Aprahamian}, A., \& {Sun}, Y. 2005, Nature Physics, 1, 81,
  \dodoi{10.1038/nphys150}

\bibitem[{{Arnould} {et~al.}(2007){Arnould}, {Goriely}, \&
  {Takahashi}}]{arnould2007review}
{Arnould}, M., {Goriely}, S., \& {Takahashi}, K. 2007, \physrep, 450, 97,
  \dodoi{10.1016/j.physrep.2007.06.002}

\bibitem[{Audi {et~al.}(2017)Audi, Kondev, Wang, Huang, \& Naimi}]{NuBase2016}
Audi, G., Kondev, F.~G., Wang, M., Huang, W.~J., \& Naimi, S. 2017, Chin. Phys.
  C, 41, 030001, \dodoi{10.1088/1674-1137/41/3/030001}

\bibitem[{Banerjee {et~al.}(2018)Banerjee, Misch, Ghorui, \&
  Sun}]{banerjee2018effective}
Banerjee, P., Misch, G.~W., Ghorui, S.~K., \& Sun, Y. 2018, Physical Review C,
  97, 065807

\bibitem[{Brown \& Rae(2014)}]{brown2014shell}
Brown, B., \& Rae, W. 2014, Nuclear Data Sheets, 120, 115

\bibitem[{Brown {et~al.}(2018)Brown, Chadwick, Capote, Kahler, Trkov, Herman,
  Sonzogni, Danon, Carlson, Dunn, Smith, Hale, Arbanas, Arcilla, Bates, Beck,
  Becker, Brown, Casperson, Conlin, Cullen, Descalle, Firestone, Gaines, Guber,
  Hawari, Holmes, Johnson, Kawano, Kiedrowski, Koning, Kopecky, Leal, Lestone,
  Lubitz, {Márquez Damián}, Mattoon, McCutchan, Mughabghab, Navratil,
  Neudecker, Nobre, Noguere, Paris, Pigni, Plompen, Pritychenko, Pronyaev,
  Roubtsov, Rochman, Romano, Schillebeeckx, Simakov, Sin, Sirakov, Sleaford,
  Sobes, Soukhovitskii, Stetcu, Talou, Thompson, {van der Marck},
  Welser-Sherrill, Wiarda, White, Wormald, Wright, Zerkle, Žerovnik, \&
  Zhu}]{ENDFB8}
Brown, D., Chadwick, M., Capote, R., {et~al.} 2018, Nuclear Data Sheets, 148, 1
  , \dodoi{https://doi.org/10.1016/j.nds.2018.02.001}

\bibitem[{{Coc} {et~al.}(1999){Coc}, {Porquet}, \& {Nowacki}}]{coc1999lt}
{Coc}, A., {Porquet}, M.-G., \& {Nowacki}, F. 1999, \prc, 61, 015801,
  \dodoi{10.1103/PhysRevC.61.015801}

\bibitem[{{Diehl} {et~al.}(1995){Diehl}, {Dupraz}, {Bennett}, {Bloemen},
  {Hermsen}, {Knoedlseder}, {Lichti}, {Morris}, {Ryan}, {Schoenfelder},
  {Steinle}, {Strong}, {Swanenburg}, {Varendorff}, \&
  {Winkler}}]{diehl1995comptel}
{Diehl}, R., {Dupraz}, C., {Bennett}, K., {et~al.} 1995, \aap, 298, 445

\bibitem[{{Dracoulis} {et~al.}(2016){Dracoulis}, {Walker}, \&
  {Kondev}}]{dracoulis2016review}
{Dracoulis}, G.~D., {Walker}, P.~M., \& {Kondev}, F.~G. 2016, Reports on
  Progress in Physics, 79, 076301, \dodoi{10.1088/0034-4885/79/7/076301}

\bibitem[{{Fujimoto} \& {Hashimoto}(2020)}]{fujimoto2020isomers}
{Fujimoto}, S.-i., \& {Hashimoto}, M.-a. 2020, \mnras, 493, L103,
  \dodoi{10.1093/mnrasl/slaa016}

\bibitem[{Gupta \& Meyer(2001)}]{gupta2001internal}
Gupta, S.~S., \& Meyer, B.~S. 2001, Physical Review C, 64, 025805

\bibitem[{{Hahn}(1921)}]{hahn1921exp}
{Hahn}, O. 1921, Naturwissenschaften, 9, 84, \dodoi{10.1007/BF01491321}

\bibitem[{{Hayakawa} {et~al.}(2009){Hayakawa}, {Shizuma}, {Chiba}, {Kajino},
  {Hatsukawa}, {Iwamoto}, {Shinohara}, \& {Harada}}]{hayakawa2009ncap}
{Hayakawa}, T., {Shizuma}, T., {Chiba}, S., {et~al.} 2009, \apj, 707, 859,
  \dodoi{10.1088/0004-637X/707/2/859}

\bibitem[{{Hayakawa} {et~al.}(2005){Hayakawa}, {Shizuma}, {Kajino}, {Chiba},
  {Shinohara}, {Nakagawa}, \& {Arima}}]{hayakawa2005sp}
{Hayakawa}, T., {Shizuma}, T., {Kajino}, T., {et~al.} 2005, \apj, 628, 533,
  \dodoi{10.1086/430198}

\bibitem[{{Horowitz} {et~al.}(2019){Horowitz}, {Arcones}, {C{\^o}t{\'e}},
  {Dillmann}, {Nazarewicz}, {Roederer}, {Schatz}, {Aprahamian}, {Atanasov},
  {Bauswein}, {Beers}, {Bliss}, {Brodeur}, {Clark}, {Frebel}, {Foucart},
  {Hansen}, {Just}, {Kankainen}, {McLaughlin}, {Kelly}, {Liddick}, {Lee},
  {Lippuner}, {Martin}, {Mendoza-Temis}, {Metzger}, {Mumpower}, {Perdikakis},
  {Pereira}, {O'Shea}, {Reifarth}, {Rogers}, {Siegel}, {Spyrou}, {Surman},
  {Tang}, {Uesaka}, \& {Wang}}]{horowitz2019review}
{Horowitz}, C.~J., {Arcones}, A., {C{\^o}t{\'e}}, B., {et~al.} 2019, Journal of
  Physics G Nuclear Physics, 46, 083001, \dodoi{10.1088/1361-6471/ab0849}

\bibitem[{{Lippuner} \& {Roberts}(2015)}]{lippuner2015heating}
{Lippuner}, J., \& {Roberts}, L.~F. 2015, \apj, 815, 82,
  \dodoi{10.1088/0004-637X/815/2/82}

\bibitem[{{Mahoney} {et~al.}(1982){Mahoney}, {Ling}, {Jacobson}, \&
  {Lingenfelter}}]{mahoney1982diffuse}
{Mahoney}, W.~A., {Ling}, J.~C., {Jacobson}, A.~S., \& {Lingenfelter}, R.~E.
  1982, \apj, 262, 742, \dodoi{10.1086/160469}

\bibitem[{Misch {et~al.}(2020)Misch, Ghorui, Banerjee, Sun, \&
  Mumpower}]{misch2020astromers}
Misch, G.~W., Ghorui, S.~K., Banerjee, P., Sun, Y., \& Mumpower, M.~R. 2020,
  The Astrophysical Journal Supplement Series, 252, 2

\bibitem[{{M{\"o}ller} {et~al.}(2019){M{\"o}ller}, {Mumpower}, {Kawano}, \&
  {Myers}}]{moller2019beta}
{M{\"o}ller}, P., {Mumpower}, M.~R., {Kawano}, T., \& {Myers}, W.~D. 2019,
  Atomic Data and Nuclear Data Tables, 125, 1,
  \dodoi{10.1016/j.adt.2018.03.003}

\bibitem[{{Mumpower} {et~al.}(2016){Mumpower}, {Surman}, {McLaughlin}, \&
  {Aprahamian}}]{mumpower2016review}
{Mumpower}, M.~R., {Surman}, R., {McLaughlin}, G.~C., \& {Aprahamian}, A. 2016,
  Progress in Particle and Nuclear Physics, 86, 86,
  \dodoi{10.1016/j.ppnp.2015.09.001}

\bibitem[{{Nesterenko} {et~al.}(2020){Nesterenko}, {Kankainen}, {Kostensalo},
  {Nobs}, {Bruce}, {Beliuskina}, {Canete}, {Eronen}, {Gamba}, {Geldhof}, {de
  Groote}, {Jokinen}, {Kurpeta}, {Moore}, {Morrison}, {Podoly{\'a}k},
  {Pohjalainen}, {Rinta-Antila}, {de Roubin}, {Rudigier}, {Suhonen},
  {Vil{\'e}n}, {Virtanen}, \& {{\"A}yst{\"o}}}]{nesterenko2020isomer}
{Nesterenko}, D.~A., {Kankainen}, A., {Kostensalo}, J., {et~al.} 2020, arXiv
  e-prints, arXiv:2005.09398.
\newblock \doarXiv{2005.09398}

\bibitem[{Okumura {et~al.}(2018)Okumura, Kawano, Jaffke, Talou, \&
  Chiba}]{okumura2018fis}
Okumura, S., Kawano, T., Jaffke, P., Talou, P., \& Chiba, S. 2018, Journal of
  Nuclear Science and Technology, 55, 1009,
  \dodoi{10.1080/00223131.2018.1467288}

\bibitem[{Orford {et~al.}(2020)Orford, Kondev, Savard, Clark, Porter, Ray,
  Buchinger, Burkey, Gorelov, Hartley, Klimes, Sharma, Valverde, \&
  Yan}]{orford2020isomer}
Orford, R., Kondev, F.~G., Savard, G., {et~al.} 2020, Phys. Rev. C, 102,
  011303, \dodoi{10.1103/PhysRevC.102.011303}

\bibitem[{Patel {et~al.}(2014)Patel, S\"oderstr\"om, Podoly\'ak, Regan, Walker,
  Watanabe, Ideguchi, Simpson, Liu, Nishimura, Wu, Xu, Browne, Doornenbal,
  Lorusso, Rice, Sinclair, Sumikama, Wu, Xu, Aoi, Baba, Bello~Garrote, Benzoni,
  Daido, Fang, Fukuda, Gey, Go, Gottardo, Inabe, Isobe, Kameda, Kobayashi,
  Kobayashi, Komatsubara, Kojouharov, Kubo, Kurz, Kuti, Li, Matsushita,
  Michimasa, Moon, Nishibata, Nishizuka, Odahara, \ifmmode~\mbox{\c{S}}\else
  \c{S}\fi{}ahin, Sakurai, Schaffner, Suzuki, Takeda, Tanaka, Taprogge, Vajta,
  Yagi, \& Yokoyama}]{patel2014midshell}
Patel, Z., S\"oderstr\"om, P.-A., Podoly\'ak, Z., {et~al.} 2014, Phys. Rev.
  Lett., 113, 262502, \dodoi{10.1103/PhysRevLett.113.262502}

\bibitem[{Raut {et~al.}(2013)Raut, Tonchev, Rusev, Tornow, Iliadis, Lugaro,
  Buntain, Goriely, Kelley, Schwengner, Banu, \& Tsoneva}]{raut2013cs}
Raut, R., Tonchev, A.~P., Rusev, G., {et~al.} 2013, Phys. Rev. Lett., 111,
  112501, \dodoi{10.1103/PhysRevLett.111.112501}

\bibitem[{Reifarth {et~al.}(2018)Reifarth, Fiebiger, G{\"o}bel, Heftrich,
  Kausch, K{\"o}ppchen, Kurtulgil, Langer, Thomas, \&
  Weigand}]{reifarth2018treatment}
Reifarth, R., Fiebiger, S., G{\"o}bel, K., {et~al.} 2018, International Journal
  of Modern Physics A, 33, 1843011

\bibitem[{Runkle {et~al.}(2001)Runkle, Champagne, \& Engel}]{runkle2001thermal}
Runkle, R., Champagne, A., \& Engel, J. 2001, The Astrophysical Journal, 556,
  970

\bibitem[{Sikorsky {et~al.}(2020)Sikorsky, Geist, Hengstler, Kempf, Gastaldo,
  Enss, Mokry, Runke, D\"ullmann, Wobrauschek, Beeks, Rosecker, Sterba,
  Kazakov, Schumm, \& Fleischmann}]{sikorsky2020th}
Sikorsky, T., Geist, J., Hengstler, D., {et~al.} 2020, Phys. Rev. Lett., 125,
  142503, \dodoi{10.1103/PhysRevLett.125.142503}

\bibitem[{Simpson {et~al.}(2014)Simpson, Gey, Jungclaus, Taprogge, Nishimura,
  Sieja, Doornenbal, Lorusso, S\"oderstr\"om, Sumikama, Xu, Baba, Browne,
  Fukuda, Inabe, Isobe, Jung, Kameda, Kim, Kim, Kojouharov, Kubo, Kurz, Kwon,
  Li, Sakurai, Schaffner, Shimizu, Suzuki, Takeda, Vajta, Watanabe, Wu, Yagi,
  Yoshinaga, B\"onig, Daugas, Drouet, Gernh\"auser, Ilieva, Kr\"oll,
  Montaner-Piz\'a, Moschner, M\"ucher, Na\"{\i}dja, Nishibata, Nowacki,
  Odahara, Orlandi, Steiger, \& Wendt}]{simpson2014yrast}
Simpson, G.~S., Gey, G., Jungclaus, A., {et~al.} 2014, Phys. Rev. Lett., 113,
  132502, \dodoi{10.1103/PhysRevLett.113.132502}

\bibitem[{{Soddy}(1917)}]{soddy1917isomer}
{Soddy}, F. 1917, The Scientific Monthly, 5, 451

\bibitem[{{Sprouse} {et~al.}(2021){Sprouse}, {Misch}, \&
  {Mumpower}}]{sprouse2021jade}
{Sprouse}, T.~M., {Misch}, G.~W., \& {Mumpower}, M.~R. 2021, arXiv preprint
  arXiv:2102.03846

\bibitem[{{Svirikhin} {et~al.}(2017){Svirikhin}, {Andreev}, {Yeremin},
  {Izosimov}, {Isaev}, {Kuznetsov}, {Kuznetsova}, {Malyshev}, {Popeko},
  {Popov}, {Sokol}, {Chelnokov}, {Chepigin}, {Schneidman}, {Gall}, {Dorvaux},
  {Brione}, {Hauschild}, {Lopez-Martenz}, {Rezynkina}, {Mullins}, {Jones}, \&
  {Mosat}}]{svirikhin2017spf}
{Svirikhin}, A.~I., {Andreev}, A.~V., {Yeremin}, A.~V., {et~al.} 2017, Physics
  of Particles and Nuclei Letters, 14, 571, \dodoi{10.1134/S1547477117040161}

\bibitem[{{Tanvir} {et~al.}(2017){Tanvir}, {Levan},
  {Gonz{\'a}lez-Fern{\'a}ndez}, {Korobkin}, {Mandel}, {Rosswog}, {Hjorth},
  {D'Avanzo}, {Fruchter}, {Fryer}, {Kangas}, {Milvang-Jensen}, {Rosetti},
  {Steeghs}, {Wollaeger}, {Cano}, {Copperwheat}, {Covino}, {D'Elia}, {de Ugarte
  Postigo}, {Evans}, {Even}, {Fairhurst}, {Figuera Jaimes}, {Fontes}, {Fujii},
  {Fynbo}, {Gompertz}, {Greiner}, {Hodosan}, {Irwin}, {Jakobsson},
  {J{\o}rgensen}, {Kann}, {Lyman}, {Malesani}, {McMahon}, {Melandri},
  {O'Brien}, {Osborne}, {Palazzi}, {Perley}, {Pian}, {Piranomonte}, {Rabus},
  {Rol}, {Rowlinson}, {Schulze}, {Sutton}, {Th{\"o}ne}, {Ulaczyk}, {Watson},
  {Wiersema}, \& {Wijers}}]{tanvir2017obs}
{Tanvir}, N.~R., {Levan}, A.~J., {Gonz{\'a}lez-Fern{\'a}ndez}, C., {et~al.}
  2017, \apjl, 848, L27, \dodoi{10.3847/2041-8213/aa90b6}

\bibitem[{{Troja} {et~al.}(2017){Troja}, {Piro}, {van Eerten}, {Wollaeger},
  {Im}, {Fox}, {Butler}, {Cenko}, {Sakamoto}, {Fryer}, {Ricci}, {Lien}, {Ryan},
  {Korobkin}, {Lee}, {Burgess}, {Lee}, {Watson}, {Choi}, {Covino}, {D'Avanzo},
  {Fontes}, {Gonz{\'a}lez}, {Khandrika}, {Kim}, {Kim}, {Lee}, {Lee}, {Kutyrev},
  {Lim}, {S{\'a}nchez-Ram{\'{\i}}rez}, {Veilleux}, {Wieringa}, \&
  {Yoon}}]{troja2017xray}
{Troja}, E., {Piro}, L., {van Eerten}, H., {et~al.} 2017, \nat, 551, 71,
  \dodoi{10.1038/nature24290}

\bibitem[{Walker \& Podoly{\'{a}}k(2020)}]{walker2020review}
Walker, P., \& Podoly{\'{a}}k, Z. 2020, Physica Scripta, 95, 044004,
  \dodoi{10.1088/1402-4896/ab635d}

\bibitem[{Walker {et~al.}(2020)Walker, Hirayama, Lane, Watanabe, Dracoulis,
  Ahmed, Brunet, Hashimoto, Ishizawa, Kondev, {et~al.}}]{walker2020properties}
Walker, P., Hirayama, Y., Lane, G., {et~al.} 2020, Physical Review Letters,
  125, 192505

\bibitem[{{Wang} {et~al.}(2020){Wang}, {N3AS Collaboration}, {Vassh}, {FIRE
  Collaboration}, {Sprouse}, {Mumpower}, {Vogt}, {Randrup}, \&
  {Surman}}]{wang2020gam}
{Wang}, X., {N3AS Collaboration}, {Vassh}, N., {et~al.} 2020, \apjl, 903, L3,
  \dodoi{10.3847/2041-8213/abbe18}

\bibitem[{Ward(1977)}]{ward1977importance}
Ward, R.~A. 1977, The Astrophysical Journal, 216, 540

\bibitem[{Watanabe {et~al.}(2014)Watanabe, Lorusso, Nishimura, Otsuka, Ogawa,
  Xu, Sumikama, S\"oderstr\"om, Doornenbal, Li, Browne, Gey, Jung, Taprogge,
  Vajta, Wu, Yagi, Baba, Benzoni, Chae, Crespi, Fukuda, Gernh\"auser, Inabe,
  Isobe, Jungclaus, Kameda, Kim, Kim, Kojouharov, Kondev, Kubo, Kurz, Kwon,
  Lane, Moon, Montaner-Piz\'a, Moschner, Naqvi, Niikura, Nishibata, Nishimura,
  Odahara, Orlandi, Patel, Podoly\'ak, Sakurai, Schaffner, Simpson, Steiger,
  Suzuki, Takeda, Wendt, \& Yoshinaga}]{watanabe2014mono}
Watanabe, H., Lorusso, G., Nishimura, S., {et~al.} 2014, Phys. Rev. Lett., 113,
  042502, \dodoi{10.1103/PhysRevLett.113.042502}

\bibitem[{{Watson} {et~al.}(2019){Watson}, {Hansen}, {Selsing}, {Koch},
  {Malesani}, {Andersen}, {Fynbo}, {Arcones}, {Bauswein}, {Covino}, {Grado},
  {Heintz}, {Hunt}, {Kouveliotou}, {Leloudas}, {Levan}, {Mazzali}, \&
  {Pian}}]{watson2019sr}
{Watson}, D., {Hansen}, C.~J., {Selsing}, J., {et~al.} 2019, \nat, 574, 497,
  \dodoi{10.1038/s41586-019-1676-3}

\bibitem[{Weisskopf \& Wigner(1930)}]{weisskopf1930calc}
Weisskopf, V., \& Wigner, E.~P. 1930, Z. Phys., 63, 54

\bibitem[{Wisshak {et~al.}(2006)Wisshak, Voss, K\"appeler, Kazakov, Be\ifmmode
  \check{c}\else \v{c}\fi{}v\'a\ifmmode~\check{r}\else \v{r}\fi{},
  Krti\ifmmode~\check{c}\else \v{c}\fi{}ka, Gallino, \&
  Pignatari}]{wisshak2006fast}
Wisshak, K., Voss, F., K\"appeler, F., {et~al.} 2006, Phys. Rev. C, 73, 045807,
  \dodoi{10.1103/PhysRevC.73.045807}

\bibitem[{Zhang {et~al.}(2019)Zhang, Watanabe, Dracoulis, Kondev, Lane, Regan,
  Söderström, Walker, Yoshida, Kanaoka, Korkulu, Lee, Liu, Nishimura, Wu,
  Yagi, Ahn, Alharbi, Baba, Browne, Bruce, Carpenter, Carroll, Chae, Chiara,
  Dombradi, Doornenbal, Estrade, Fukuda, Griffin, Ideguchi, Inabe, Isobe,
  Kanaya, Kojouharov, Kubo, Kubono, Kurz, Kuti, Lalkovski, Lauritsen, Lee, Lee,
  Lister, Lorusso, Lotay, McCutchan, Moon, Nishizuka, Nita, Odahara, Patel,
  Phong, Podolyák, Roberts, Sakurai, Schaffner, Seweryniak, Shand, Shimizu,
  Sumikama, Suzuki, Takeda, Terashima, Vajta, Valiente-Dóbon, Xu, \&
  Zhu}]{zhang2019isomer}
Zhang, G., Watanabe, H., Dracoulis, G., {et~al.} 2019, Physics Letters B, 799,
  135036, \dodoi{https://doi.org/10.1016/j.physletb.2019.135036}

\bibitem[{{Zhu} {et~al.}(2018){Zhu}, {Wollaeger}, {Vassh}, {Surman}, {Sprouse},
  {Mumpower}, {M{\"o}ller}, {McLaughlin}, {Korobkin}, {Kawano}, {Jaffke},
  {Holmbeck}, {Fryer}, {Even}, {Couture}, \& {Barnes}}]{zhu2018cf}
{Zhu}, Y., {Wollaeger}, R.~T., {Vassh}, N., {et~al.} 2018, \apjl, 863, L23,
  \dodoi{10.3847/2041-8213/aad5de}

\end{thebibliography}
\bibliographystyle{aasjournal}

\end{document}